# Agentic workflow enables the recovery of critical materials from complex feedstocks via selective precipitation

Andrew Ritchhart,[a,#] Sarah I. Allec,[a,#] Pravalika Butreddy,[a] Krista Kulesa,[a] Qingpu Wang,[b] Dan Thien Nguyen,[a] Maxim Ziatdinov,[a,*] Elias Nakouzi[a,c,*]

[a.] Physical and Computational Sciences Directorate, Pacific Northwest National Laboratory, Richland, Washington 99354, United States.
[b.] Physical and Computational Sciences Directorate, Pacific Northwest National Laboratory, Seattle, Washington 98109, United States.
[c.] Department of Chemical Engineering, University of Washington, Seattle, Washington 98195, United States.
*maxim.ziatdinov@pnnl.gov, *elias.nakouzi@pnnl.gov

**Abstract.** We present a multi-agentic workflow for critical materials recovery that deploys a series of AI agents and automated instruments to recover critical materials from produced water and magnet leachates. This approach achieves selective precipitation from real-world feedstocks using simple chemicals, accelerating the development of efficient, adaptable, and scalable separations to a timeline of days, rather than months and years.

The recovery of critical materials from complex feedstocks is a problem of increasing urgency due to the surging demand for these elements and their importance to energy technologies.[1] One of the cornerstone techniques for critical materials separation is selective precipitation, wherein a chemical reagent added to the mixture reacts with the less soluble ions, creating solid precipitates that can subsequently be recovered. Precipitation is widely applied in the separations industry due to its scalability, chemical versatility, and low energy expense.[2] Recent studies have extended this approach to unconventional feedstocks, including polymetallic nodules, recycled electronics, seawater, low-grade ores, and more.[3-10] By finely tuning the chemical conditions and controlling ion speciation, precipitation can outcompete separation approaches that rely on tailored and engineered materials such as membranes, metal organic frameworks, ionic liquids, and ligands,[11,12] especially when factoring energy-efficiency and economic considerations.

One key limitation of selective precipitation is that the products often incorporate significant amounts of impurities, resulting in inclusions or solid solutions that require further purification. This challenge is compounded by the nature of real-world feedstocks, which are typically highly variable and complex, often including multiple competing ions with similar physicochemical properties.[13] Predicting the composition of the precipitates from complex feedstocks and thus the separation efficiency remains beyond the limits of physics-based speciation and crystal growth models. As such, optimizing precipitation-based separations across this vast parameter space presents an excellent use case for artificial intelligence (AI)-powered workflows and autonomous experimentation.

Recent advances in artificial intelligence,[14-18] particularly large language models (LLMs),[19-21] have enhanced and widened the possibilities for AI-driven scientific experimentation. However, effectively deploying these capabilities in laboratory settings remains challenging. Stand-alone LLMs lack the specialization and niche domain knowledge required for many areas of scientific research. Conversely, platforms built specifically for laboratory automation – including LLM-enabled systems – often suffer from being overly rigid, demanding extensive integration efforts and programming expertise to customize and deploy. Truly transformative AI-driven scientific innovation will require platforms that are modular and adaptive yet resource-efficient, while remaining accessible to domain scientists. Agentic AI – systems in which LLM-based agents can autonomously plan, execute, and iteratively refine multi-step workflows – offers a promising path forward by combining the flexible reasoning of LLMs with the ability to dynamically invoke specialized tools, instruments, and domain knowledge as needed, all primarily through natural language interaction.

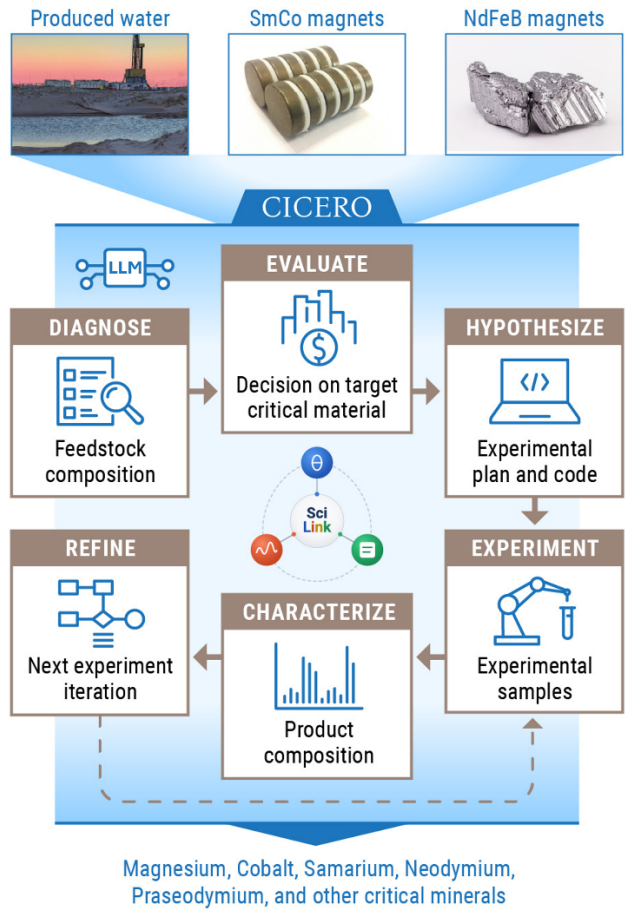

**Figure 1. Description of CICERO workflow, powered by multi-agentic SciLink platform**. CICERO takes as input a real-world feedstock composition, and provides as output an optimized separation method to recover critical materials via selective precipitation.

In realizing this vision, one of the biggest challenges is defining scientific problems that are simultaneously ambitious enough to merit the development of agentic workflows, while being tractable with current AI agents, lab instrumentation, and reasonable constraints on the science questions being asked. In this context, adapting selective precipitation using simple commodity chemicals and existing liquid handling automation is an impactful and attainable goal. Here, we demonstrate an agentic AI-guided closed-loop workflow that autonomously plans, evaluates, and optimizes selective precipitation campaigns for critical materials recovery from complex feedstocks and waste streams.

Our workflow is built on SciLink,[22] a recently developed LLM-powered multi-agent platform that automates experimental design, data analysis, and computational modelling tasks in materials research. We apply its planning subsystem – which guides researchers through the full campaign loop from literature-grounded hypothesis generation and economic viability screening, through automated data analysis, to closed-loop Bayesian optimization – to critical materials recovery, a workflow we refer to as CICERO (Computer Intelligence for Critical Elements Recovery and Optimization). Within CICERO, the planning subsystem manages the full lifecycle of an experimental campaign through three coordinated SciLink sub-agents. The Planning Agent ingests domain literature (papers, technical reports) and any existing experimental data into a retrieval-augmented generation pipeline, producing testable hypotheses, detailed experimental protocols, and economic viability screenings – all grounded in and traceable to the source documents. Critically, the viability screening step acts as an economic filter that narrows the hypothesis and planning space before any experiments begin, steering the campaign toward materials, processes, and parameter ranges worth pursuing given real-world cost drivers, market viability, and scalability constraints. Once experiments are underway, raw instrument data is converted into scalar descriptors suitable for mathematical optimization. While SciLink includes a Scalarizer Agent to automate this task by auto-generating analysis scripts, the data conversion in this work was straightforward, and we employed a pre-existing, validated script. These descriptors then feed into the Bayesian optimization (BO) Agent, which drives a Bayesian optimization loop[23] – supporting single- and multi-objective campaigns, budget-aware acquisition strategies, batch parallelism for high-throughput platforms, and physical constraint handling for real-world setups such as multi-well plates.

We tested CICERO by processing three types of feedstocks: 1) produced water from oil and gas extraction in Oklahoma, 2) commercially sourced samarium–cobalt (SmCo) permanent magnets, and 3) commercially sourced neodymium–iron–boron (NdFeB) permanent magnets. Details on the feedstock compositions and pre-processing steps, such as magnet acid leaching, are provided in the SI. For each feedstock, the end point after processing the sample with CICERO is a well-defined, feedstock-specific protocol for the recovery of critical materials using simple commodity chemicals. The goal of this work is not to develop new separation chemistries and technologies, but rather to rapidly identify and optimize separation methodologies that are feedstock-specific, economically feasible, and readily scalable.

In the following, we provide additional details on the sequential steps, AI agents, and current roles of humans in CICERO. 1) The initial "Diagnose" step includes characterization of the feedstock chemistry, providing the basis data for attempting hundreds of separation experiments in the subsequent steps. Specifically, quantitative Inductively Coupled Plasma Mass Spectrometry (ICP-MS) measurements produce a table of the identity and concentrations of the elements present in the feedstock. The SciLink Planning Agent ingests this literature and data to map the problem space. 2) Next is the "Evaluate" step, where the SciLink Planning Agent determines the target critical materials in the feedstock based on a preliminary technoeconomic analysis of value, concentration, criticality, and anticipated range of product purity. The agent is optionally augmented at run time with additional data, for example in our described use case including US DOE Critical Materials Assessment documents. 3) The decision on the target critical materials is passed onto the "Hypothesize" step, where the SciLink Planning Agent generates one or more testable pathways for recovering the target critical material. By prompting, we specifically target in this work a series of precipitation and dissolution steps; an approach that remains the backbone of the chemical separations industry. To this end, the agent is constrained to using commodity chemicals, such as simple acids and bases, as well as precipitants, such as oxalic acid, citric acid, sodium bicarbonate, or hydrogen peroxide.

The next steps involve physical experiments: 4) In the "Experiment" step, the agent converts the hypothesis into an initial batch of 96 experiments that span the relevant parameter space, including the identity and concentration of the chemicals used for separation, as well as the sequence and time of adding these chemicals. The agent then provides an actionable Python script that automatically executes the experiments in a 96-well plate using an Opentrons robot. After the experiments are complete, a human operator moves the samples between the Opentrons robot and a microplate centrifuge for multiple cycles of washing and decanting, before using automatically generated code to dissolve the purified products in preparation for ICP-MS measurements. A human then moves the plate to the ICP-MS instrument, which automatically measures the chemical composition in the products of all the samples. 5) The results are then passed to a pre-existing script for processing the ICP-MS data files and resolving the composition of the products across the parameter space. 6) The final "Refine" step involves determining whether satisfactory separations have been achieved by evaluating the % purity by mol relative to the other metals and % yield of the target critical material. This involves the optional use of the Bayesian Optimization agent or re-prompting the planning agent to recommend additional iterations on experiments, closing the loop and converging towards an optimized separations method. CICERO is programmed to offer a choice of either fully automated analysis and planning or human-in-the-loop feedback.

In the experiments described, human feedback was kept minimal and used at three points: in the experimental agent to modify maximum allowable reagent concentrations, in code generation to supervise and fix any minor bugs, and in refinement to inform the target purity metric of subsequent steps. All agents utilized Google's gemini-2.5-promodel.[24] Verbatim SciLink agent input and output text are available in the SI.

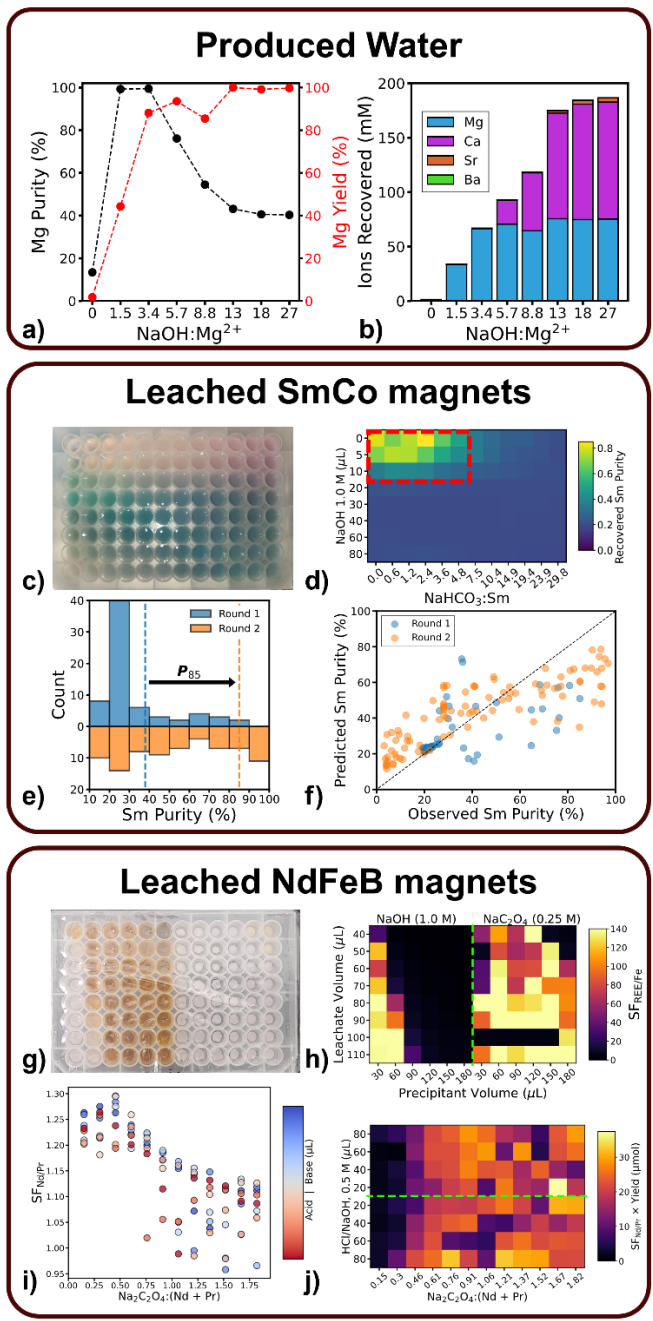

**Figure 2. Experimental results of critical materials separations from three real-world feedstocks, produced water, leached SmCo magnets, and leached NdFeB magnets. a.** Purity in mol% (black) and corresponding yield (red) of Mg recovered as precipitate via addition of NaOH solution to produced water. **b**. Recovered Mg, Ca, Sr, and Ba from produced water as precipitates digested for ICP-MS measurements. **c**. Image of plate containing SmCo feedstock after Round 1 of precipitation with $NaHCO_3$ and NaOH. Sm salts are colorless to faint yellow whereas Co salt precipitates range from pink to green and blue depending on crystalline phase of the hydroxide formed. **d**. Purity of Sm salt recovered from the SmCo feedstock after the first round of precipitation experiments (where a value of 1 represents 100% purity relative to other metals). **e.** Histogram of purities recovered from the first (blue) and second (orange) rounds of 96 precipitation experiments of SmCo feedstock. Dashed lines correspond to the 85th percentile of

recovery for corresponding rounds. **f.** Parity plot of the Bayesian Optimization machine learning on Sm % purity as a function of added NaHCO$_3$ and NaOH after the two rounds of data collection. **g**. Image of a plate containing NdFeB feedstock after precipitation with NaOH (left) and Na$_2$C$_2$O$_4$ (right). REE precipitates are pale to colorless while Fe precipitates are orange. **h**. Separation factor of Fe relative to combined REE total after Round 1 of precipitation of the NdFeB feedstock. **i.** The separation factor of Nd from Pr after Round 2 of precipitation of NdFeB stock after Fe removal, as a function of added Na$_2$C$_2$O$_4$ and NaOH (blue) and HCl (red). j. The separation factor from Figure multiplied by the yield.

Using CICERO, we processed three types of real-world feedstocks, deploying distinct approaches for critical materials separations. For the produced water feedstock, CICERO identified magnesium as the highest value target. Despite noting the presence of other more valuable components, such as copper on a per-atom basis, magnesium was selected due to its exceedingly high concentration (1,840 ppm) and cost effectiveness for recovery. The agent-generated hypothesis was tuning the pH to alkaline conditions to selectively precipitate magnesium hydroxide, bypassing the other alkaline earth metals present in substantial amounts, including calcium (>13,000 ppm), strontium (450 ppm), and barium (2 ppm), as well as more soluble ions such as sodium (>6,000 ppm). This approach involved the addition of a single precipitating agent, sodium hydroxide, in various concentrations and evaluating the composition of the solid precipitates. The agent opted for a molar ratio of NaOH to Mg between 1.5:1 and 2:1, targeting 8 experimental conditions with 12 sample duplicates. We tested the reproducibility of robotic experiments, the iterative centrifugation and rinsing, and the automated ICP-MS measurements, which produced ion concentration measurements within a reasonable standard deviation of 4.0–7.6% and 2 outlier values from the batch of 96 samples.

Within a single iteration of these 96 experiments, 99.4% pure magnesium hydroxide salt was recovered at 86% yield upon addition of 3.4 equivalent of NaOH (Figure 2a). Lower [NaOH] also produced high-purity magnesium hydroxide, but with sub-optimal yields. Conversely, higher equivalents and concentrations of NaOH resulted in the inclusion of the competing alkaline earth metals, eventually forming precipitate mixtures that are 41.4% Mg, 56.5% Ca, 2.0% Sr, and 0.007% Ba calculated as mol % relative to all the metals in the product, reflecting the composition of the metal ions in the feedstock (Figure 2b). To demonstrate the scalability of these results, we performed bulk precipitation experiments using a thousand-fold larger volumes, reporting similar purity and yield of the magnesium product (details in SI, Figure S2).

For the second feedstock, leached commercial SmCo magnets, the SciLink's Planning Agent identified both samarium and cobalt as valuable candidates for extraction, with samarium being the primary target. Unlike the produced water feedstock, here the recommended hypothesis involved two reagents, specifically adding both sodium bicarbonate and sodium hydroxide in various amounts. The underlying hypothesis was using a pH of approximately 6.5 to alter the speciation of cobalt complexes, while selectively precipitating samarium using the carbonate reagent. Such carbonate selectivity towards rare earth elements notably has literature precedent[25]. The SciLink's Planning Agent converted this hypothesis into a range of 96 experimental conditions and generated the corresponding Opentrons python script (SI). The parameter space spanned a range of [NaHCO$_3$] = 5–200 mM and [NaOH] = 12.5–250 mM. The experiments resulted in products with a variety of colors; pink at high [NaHCO$_3$] and low [NaOH], light brown at the opposite extreme of high [NaOH] and low [(NH$_4$)$_2$CO$_3$], and a gradient of bluish green at most other conditions (Figure 2c). These variations likely correspond to cobalt (oxy)hydroxide polymorphs formed at different solution conditions and were correlated with changes in the UV-vis spectra (Figure S4). ICP-MS analysis showed reasonable separation after one round, achieving an 85.0% Sm salt product from an initially 21.9% Sm feedstock relative to the total metal ion composition, corresponding to a separation factor of 22.8, defined as $SF_{Sm/Co} = \frac{n_{Sm,f}/n_{Sm,i}}{n_{Co,f}/n_{Co,i}}$ (Figure 2d). To further improve this outcome, we performed a second batch of experiments employing Sci-Link's built-in Bayesian optimization agent in conjunction with the planning agent. The second-round experimental plan and corresponding code suggested the design space of [NaOH] = 0.27-0.675 mM and [NaHCO$_3$] = 20–80 mM, resulting in an increased purity of Sm product with a shift in the 85$^{th}$ percentile from ~40% in the first batch of experiments to 89% at 92>99 % yield (Figure 2e, f), with noticeably improved confidence intervals in the high purity region.

The third feedstock that we investigated was the NdFeB magnet leachate, prompting yet a different approach from CICERO, which opted to explore two precipitating reagents separately, namely sodium hydroxide and sodium oxalate. The Planning Agent identified neodymium (23,800 ppm) and praseodymium (5,900 ppm) as the primary targets, as well as the lower concentrations of dysprosium (27 ppm), terbium (16 ppm), and cobalt (470 ppm) as potential co-products, and attempted to bypass the substantial amounts of iron (62,000 ppm) and boron (920 ppm) in the feedstock. Our analysis shows that the oxalate was highly selective in recovering >99% pure rare earth element salts including both neodymium and praseodymium salts at 93% yield, reaching separation factors of approximately 200 relative to iron across a wide concentration regime (Figure 2g, h). By comparison, hydroxide achieved reasonable separation factors at low precipitant concentrations, likely forming double sulfate salts as the magnet was leached in sulfuric acid,[26] but the yield was only 26%. At higher concentration, hydroxide precipitates incorporated iron in large amounts significantly compromising the purity. Since the removal of iron was achieved in the first round of oxalate experiments, no further Bayesian optimization was required. Instead, we opted to prompt the Planning Agent in the second round to improve the separation factor among the two rare earth elements, Nd and Pr. The agent suggested the continued use of oxalate with pH adjustment as a second tuning parameter. A single round of this strategy yielded a Nd/Pr separation factor of up to 1.35 with a strong dependence on oxalate concentration and weak dependence on pH (Figure 2i, j). This separation factor is competitive for rare earth element separations in a single round, especially since Nd and Pr are adjacent elements in the periodic table and notoriously challenging to separate. The conditions were explored nearly autonomously using the workflow described earlier, over a period of approximately three days.

In conclusion, we have demonstrated a multi-agentic workflow that deploys a series of agents and automated instruments to achieve the separation and recovery of critical materials from a variety of complex feedstocks. In the current workflow, the role of "humans-in-the-loop" is restricted to two categories of actions. The first is trivial tasks, such as prompting agents with preliminary data, clicking on instrument software buttons to conduct experiments, and moving the sample plates between the instruments. These tasks do not have significant bearing on the scientific nature of the workflow and current efforts to fully automate them are in progress. The other category of human intervention involves supervising the output from key science agents that perform the hypothesis-generation, experiment generation, and data processing. This consideration is even more relevant when applying the workflow to problems that push the boundaries of the literature corpus, such as materials discovery, rather than the optimization of relatively simple chemistries as described here. In the current work, intervention was kept to a bare minimum to better reflect SciLink's capabilities; examples of this included correcting CICERO on the maximum concentration of sodium oxalate, adding a trash bin location to the Opentrons script, and suggesting working volumes be kept above 5 µL for precision.

Building on the current version of the workflow, we anticipate that the "Evaluate" and "Hypothesize" steps could be incorporated in the iterative loop after the results are analysed, depending on how the experimentation is progressing for two reasons: Firstly, this iteration mitigates the risk of an initially bad hypothesis producing a narrow parameter space that is sub-optimal and challenging for the agents to reason out of. Moreover, this could help identify opportunities for recovering additional critical materials that were not initially designated as the targets, but whose recovery becomes viable as new experimental results are processed. Substantial research efforts continue to focus on achieving and optimizing critical materials separations via selective precipitation; CICERO demonstrates how the efficiency of this research can be massively enhanced using agentic workflows and automated experiments.

Lastly, CICERO's ability to rapidly pivot between fundamentally different separation strategies for each feedstock demonstrates the adaptability that makes agentic approaches particularly valuable when processing diverse, real-world feedstocks. Furthermore, the data generated by CICERO – with full experimental provenance from hypothesis to execution to analysis and decision-making – creates a reproducible knowledge base that can inform future separations research. As global demand for critical materials intensifies and feedstock complexity increases, agentic workflows like CICERO offer a scalable path toward rapidly developing feedstock-specific, economically viable separation processes that would be impractical to optimize through traditional approaches alone.

**Acknowledgments**

Feedstock acquisition and separation experiments were supported by the Non-Equilibrium Transport Driven Separations (NETS) initiative, under the Laboratory Directed Research and Development (LDRD) Program at Pacific Northwest National Laboratory (PNNL). Data processing and management was supported by the Adaptive Tunability for Synthesis and Control via Autonomous Learning on Edge (ATSCALE) initiative, under the LDRD program at PNNL. The early development of SciLink was supported by seed LDRD funding from the Physical and Computational Sciences Directorate at PNNL. The setup of the automated instruments was supported by the Foundational Autonomy Investment (FAI), under the LDRD program at PNNL. PNNL is a multi-program national laboratory operated for the U.S. Department of Energy (DOE) by Battelle Memorial Institute under Contract No. DE-AC05-76RL01830. The graphic in Figure 1 was developed by Nathan Johnson.

**Conflicts of interest**

There are no conflicts to declare.

**References**

(1) Sholl, D. S.; Lively, R. P. Seven chemical separations to change the world. *Nature* **2016**, *532*, 435–437.

(2) National Academies of Sciences, Engineering, and Medicine: A Research Agenda for Transforming Separation Science. The National Academies Press. : Washington, DC, 2019.
(3) Yang, Y.; Walton, A.; Sheridan, R.; Güth, K.; Gauß, R.; Gutfleisch, O.; Buchert, M.; Steenari, B.-M.; Van Gerven, T.; Jones, P. T.; et al. REE Recovery from End-of-Life NdFeB Permanent Magnet Scrap: A Critical Review. *Journal of Sustainable Metallurgy* **2016**, *3* (1), 122–149. DOI: 10.1007/s40831-016-0090-4.
(4) Wang, Q.; Nakouzi, E.; Ryan, E. A.; Subban, C. V. Flow-Assisted Selective Mineral Extraction from Seawater. *Environ. Sci. Technol. Lett.* **2022**, *9*, 645–659.
(5) Wang, Q.; Fu, Y.; Miller, E. A.; Song, D.; Brahana, P. J.; Ritchhart, A.; Xu, Z.; Johnson, G. E.; Bharti, B.; Sushko, M. L.; et al. Selective Recovery of Critical Minerals from Simulated Electronic Wastes via Reaction-Diffusion Coupling. *ChemSusChem* **2025**, e202402372. DOI: doi.org/10.1002/cssc.202402372.
(6) Butreddy, P.; Mergelsberg, S. T.; Jocz, J. N.; Li, D.; Prabhakaran, V.; Ritchhart, A. J.; Subban, C. V.; Kellar, J.; Beeler, S. R.; Keenan, S. W.; et al. Selective dissolution and re-precipitation by pH cycling enables recovery of manganese from surface nodules. *RSC Sustainability* **2025**. DOI: 10.1039/d4su00444b.
(7) Shekarian, Y.; Rezaee, M.; Pisupati, S. Green chemical precipitation of manganese, cobalt, and nickel from acid mine drainage using ozone: mechanism and chemical kinetics. *Reaction Chemistry & Engineering* **2025**, *10* (10), 2398–2411. DOI: 10.1039/d5re00222b.

(8) Zhang, X.; Zeng, L.; Wang, Y.; Tian, J.; Wang, J.; Sun, W.; Han, H.; Yang, Y. Selective separation of metals from wastewater using sulfide precipitation: A critical review in agents, operational factors and particle aggregation. *J Environ Manage* **2023**, *344*, 118462. DOI: 10.1016/j.jenvman.2023.118462 From NLM Publisher.
(9) Li, Q.; Xiao, Z.; Zhang, W. Sulfide precipitation characteristics of Mn, Ni, Co, and Zn in the presence of contaminant metal ions. *Minerals Engineering* **2024**, *215*. DOI: 10.1016/j.mineng.2024.108814.
(10) Sun, P.; Jing, D.; Durham, D. B.; Sapkota, B.; Chen, Y.; Anderson, J. L. Ion-Specific Precipitation of Extractants Enables Rare-Earth Separation and Wastewater Remediation from Solvent Extraction of Critical Elements. *Environ Sci Technol* **2026**. DOI: 10.1021/acs.est.5c16966 From NLM Publisher.
(11) Xu, L.; Zhao, B.; Zhang, X.; Liu, W.; Rau, D.; Wu, D.; Zhang, W.; Liu, C.; Liu, Z.; Lin, S. Membrane and electrochemical separations for direct lithium extraction. *Nature Chemical Engineering* **2025**, *2*, 551–567.
(12) Kazi, O. A.; Chen, W.; Eatman, J. G.; Gao, F.; Liu, Y.; Wang, Y.; Xia, Z.; Darling, S. B. Material Design Strategies for Recovery of Critical Resources from Water. *Adv Mater* **2023**, *35* (36), e2300913. DOI: 10.1002/adma.202300913 From NLM PubMed-not-MEDLINE.
(13) Can Sener, S. E.; Thomas, V. M.; Hogan, D. E.; Maier, R. M.; Carbajales-Dale, M.; Barton, M. D.; Karanfil, T.; Crittenden, J. C.; Amy, G. L. Recovery of Critical Metals from Aqueous Sources. *ACS Sustain Chem Eng* **2021**, *9* (35), 11616–11634. DOI: 10.1021/acssuschemeng.1c03005 From NLM PubMed-not-MEDLINE.
(14) Vaswani, A.; Shazeer, N.; Parmar, N.; Uszkoreit, J.; Jones, L.; Gomez, A. N.; Kaiser, L.; Polosukhin, I. Attention is All you Need. In *NeurIPS*, 2017.
(15) Radford, A.; Wu, J.; Child, R.; Luan, D.; Amodei, D.; Sutskever, I. Language Models are Unsupervised Multitask Learners. *Semantic Scholar* **2019**, CorpusID:160025533.
(16) Christiano, P. F.; Leike, J.; Brown, T. B.; Martic, M.; Legg, S.; Amodei, D. Deep reinforcement learning from human preferences. In *NeurIPS*, 2017.
(17) Devlin, J.; Chang, M.-W.; Lee, K.; Toutanova, K. BERT: Pre-training of Deep Bidirectional Transformers for Language Understanding. In *NAACL-HLT*, 2019; pp 4171–4186.
(18) Radford, A.; Narasimhan, K. Improving Language Understanding by Generative Pre-Training. *Semantic Scholar* **2018**, CorpusID:49313245.
(19) Brown, T. M., B.; Ryder, N.; Subbiah, M.; Kaplan, J. D.; Dhariwal, P.; Neelakantan, A.; Shyam, P.; Sastry, G.; Askell, A.; Agarwal, A.; Herbert-Voss, A.; Krueger, G.; Henighan, T.; Child, R.; Ramesh, A.; Ziegler, D.; Wu, J.; Winter, C.; Hesse, C.; Chen, M.; Sigler, E.; Litwin, M.; Gray, S.; Chess, B.; Clark, J.; Berner, C.; McCandlish, S.; Radford, A.; Sutskever, I.; Amodei, D. Language Models are Few-Shot Learners. In *NeurIPS*, 2020.
(20) Wei, J.; Wang, X.; Schuurmans, D.; Bosma, M.; Ichter, B.; Xia, F.; Chi, E. H.; Le, Q. V.; Zhou, D. Chain-of-thought prompting elicits reasoning in large language models. In *NeurIPS*, 2022.
(21) Ouyang, L.; Wu, J.; Jiang, X.; Almeida, D.; Wainwright, C.; Mishkin, P.; Zhang, C.; Agarwal, S.; Slama, K.; Ray, A.; et al. Training language models to follow instructions with human feedback. In *NeurIPS*, 2022.
(22) GitHub URL: https://github.com/ziatdinovmax/SciLink

(23) Shahriari, B.; Swersky, K.; Wang, Z.; Adams, R. P.; de Freitas, N. Taking the Human Out of the Loop: A Review of Bayesian Optimization. *Proceedings of the IEEE* **2015**, *104*, 148–175.
(24) al., A. e. Gemini: A Family of Highly Capable Multimodal Models. *arXiv:2312.11805* **2023**.
(25) Ren, X.; Wei, L.; Sun, A.; Xu, B.; Jiang, W.; Yang, B.; Wang, F. Recycling of Sm-Co permanent magnet waste: Perspectives and recent advances. *Journal of Rare Earths* **2025**. DOI: 10.1016/j.jre.2025.09.038.
(26) Lyman, J. W.; Palmer, G. R. Recycling of Rare Earths and Iron from NdFeB Magnet Scrap. *High Temperature Materials and Processes* **1993**, *11*, 175–187.

# Supplementary Information: Agentic workflow enables the recovery of critical materials from complex feedstocks via selective precipitation

Andrew Ritchhart,[a,#] Sarah I. Allec,[a,#] Pravalika Butreddy,[a] Krista Kulesa,[a] Qingpu Wang,[b] Dan Thien Nguyen,[a] Maxim Ziatdinov,[a,*] Elias Nakouzi[a,c,*]

## SciLink Agent Descriptions

**Planning agent.** SciLink's `PlanningAgent` is a stateful AI agent for experimental planning and iterative refinement in scientific workflows. The agent combines retrieval-augmented generation (RAG), multimodal reasoning, and human-in-the-loop feedback to generate executable experimental protocols from high-level research objectives. In particular, it uses a *dual knowledge base* for RAG in order to partition *i*) scientific knowledge, such as the scientific literature, user manuals, and reports, from *ii*) code-based knowledge, such as API documentation and code repositories. This partitioning enables efficient chunking for these two types of knowledge. Both knowledge bases use Facebook AI Similarity Search (FAISS)[1] indices for efficient similarity search and support incremental updates to avoid redundant reprocessing.

In order to capture, persist, and manage context within and across interactions, all agent actions are tracked in a *persistent state dictionary* containing:

- Objective and iteration history
- Experimental plans with scientific justification
- Generated implementation code
- Human feedback log
- Experimental results from previous iterations

All planning agent outputs are available in the SI as json files (see `*Experiment_Agent.json` and `*TEA_Agent_Analysis.json` in `Agent_Outputs/`).

**Bayesian optimization agent.** SciLink's `BOAgent` is a stateful AI agent for Gaussian-process-based Bayesian optimization (BO) with LLM-driven strategy configuration. Unlike traditional BO frameworks that require upfront hyperparameter selection and intensive model evaluation, the `BOAgent` uses vision-language models to inspect diagnostic plots and dynamically adjust acquisition strategies (*e.g.*, switching from exploration to exploitation as data accumulates).

The agent is designed for *stop-and-go* experimental campaigns with human-in-the-loop: researchers run the agent, receive recommendations, perform experiments, and restart the agent. To allow users to shut down between optimization cycles and seamlessly resume the optimization loop where they left off, the agent maintains persistent state across experimental campaigns through checkpoint files. All context is reconstructed at each invocation by reading:

- The current data file (`.xlsx` or `.csv`), which contains all experimental results to date
- A state json file, which records the optimization trajectory, including past configurations, recommendations, budget context, and acquisition function metadata

This design accommodates realistic experimental timelines where data acquisition is the bottleneck and ensures that shutting down the process (intentionally or due to system failure) does not cause state loss. Each time the agent runs, it:

- Loads the complete experimental dataset from disk
- Reads the optimization history to understand past decisions and strategy evolution
- Computes any derived context needed for the current iteration (e.g., budget remaining, data points seen, current best parameters)
- Selects a strategy, fits the Gaussian process model, and recommends the next experiments
- Appends the new iteration to the history log on disk

An example is available in `Agent_Outputs/SmCo_Round2_BO_Agent.state.json`.

## Separation and Characterization Methods

### Opentrons System

Experiments were performed on the Opentrons platform using a Flex model Opentrons liquid handling robot equipped with an XYZ lateral-motion gripper as well as P1000 single- and 8-channel pipettes (5-1000 µL). Every liquid-handling experiment was equipped with new Opentrons pipet tips, NEST liquid reservoirs, and microtiter well plates. All disposable labware is ANSI/SLAS-format. All Opentrons manipulations were programmed and edited through python-based protocols. Before running protocols, all disposable deck elements were manually calibrated to labware-specific positional offsets. Protocols retrieved polypropylene non-filtered 1000-, 200- and 50-µL pipet tips from Opentrons tip racks. Experimental solutions originated from 15- to 195-mL single- and 12-well NEST reservoirs. Solutions were transferred into and further processed in Caplugs Evergreen polypropylene flat-bottom 96-well plates (0.37 mL) and Agilent InfinityLab polypropylene round-bottom 96-well plates (0.5 mL, 1.2 mL, 2.0 mL). Microtiter samples were heated and/or agitated by the Opentrons Heater-Shaker Module up 95 °C and 3000 rpm.

### ICP-MS Elemental Analysis

High-throughput ICP-MS of samples prepared via Opentrons experiments were measured using an Agilent 8900 QQQ instrument. Samples were centrifuged in an Eppendorf 5810 R 15-amp centrifuge for 10 min at 4500 RPM and room temperature, then subsequently digested in 400 μL of 3 M trace metal grade nitric acid, covered, and incubated at 60 °C overnight. Samples were diluted in 2% acid matrix and data were acquired in He option gas mode.

**UV-Vis Spectroscopy**

High-throughput UV-Vis were collected using a Thermofisher Varioskan LUX. Digested SmCo samples were measured around the $Co^{2+}$ transition from 460 to 610 nm using 1 nm slit width and 200 ms dwell time.

## Feedstock Preparation Methods

**Samarium Cobalt magnet leachate generation**

Grade 18 Samarium Cobalt disc magnets ($SmCo_5$) were purchased from Magnetshop (Product No. 18DRE1304). For de-magnetization, the magnet discs were vacuum-sealed in a quartz ampoule and heated to 700 °C at a ramp rate of 10 °C/min. The temperature was held constant for 2 hours in a tube furnace, followed by natural cooling to room temperature. Afterward, the magnet discs were manually crushed into fine powder using an agate mortar and pestle for approximately 10 minutes to ensure uniform particle size.

The leaching solution was prepared by diluting concentrated sulfuric acid (98%, Sigma-Aldrich) to obtain a 2 M $H_2SO_4$ solution. A total of 4 g of $SmCo_5$ magnet powder was added to 200 mL of the prepared 2M $H_2SO_4$ solution. The mixture was stirred continuously at a constant rate of 200 rpm for 72 hours at room temperature to facilitate dissolution. After the reaction period, the solution was extracted from the remaining solid residues for further analysis. The composition of the leaching solution was determined using ICP-MS.

**Neodymium magnet leachate generation**

NdFeB magnets (grade N42, 0.25"× 0.25"× 0.25" cubes, part number B444; K&J Magnetics) were processed to generate leachate using previously reported hydrometallurgical methods.[2,3] The magnets were first demagnetized by heating at 650°C for 2 h in a box furnace (FD1545M, Thermolyne). The demagnetized material was then ground in a tungsten carbide-lined vial using a ball mill (BM-400-115, Cole-Parmer) for 1 h. The resulting magnet powder was leached in 2.0 M $H_2SO_4$ at a pulp density of 100 g/L. The slurry was stirred for 3 days at room temperature and subsequently centrifuged to collect the leachate.

For elemental analysis, leachate samples were digested in 70% $HNO_3$ overnight, diluted to final concentrations of 10 – 500 ppb, and analyzed by inductively coupled plasma mass spectrometry (ICP-MS, Thermo Scientific iCAP Q and Elemental Scientific seaFAST autosampler).

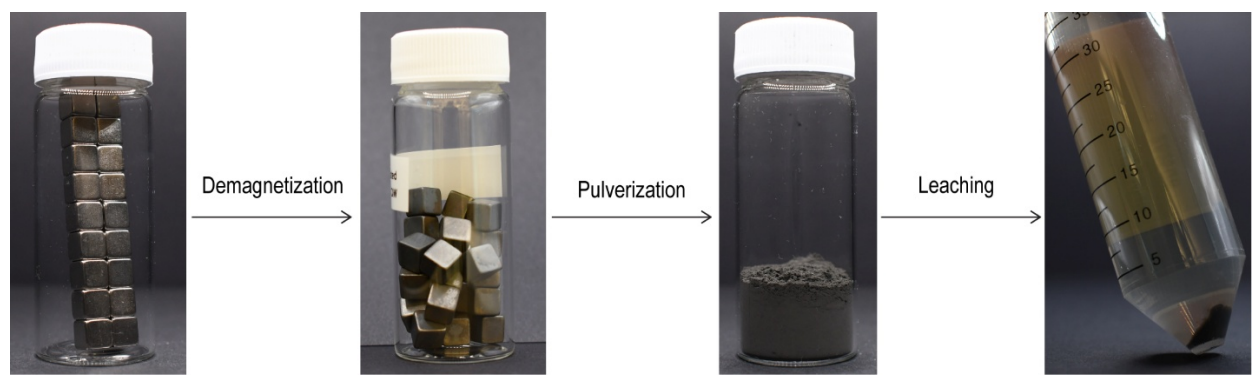

**Fig. S1.** Photographs of NdFeB magnet samples at different steps of the leachate generation process.

**Fig S2. Produced water bulk experiments**

We aimed to scale up the recovery of magnesium from produced water using bulk precipitation with NaOH as recommended by CICERO. Direct magnesium recovery was achieved at high alkalinity; the addition of 9 mL 0.5 M NaOH into 20 mL of the produced water increased the pH to 12.07, and we were able to recover ~0.11g of pure magnesium-rich precipitates (98.68 mol% Mg, 0.62 mol% Na, and 0.69 mol% Ca).

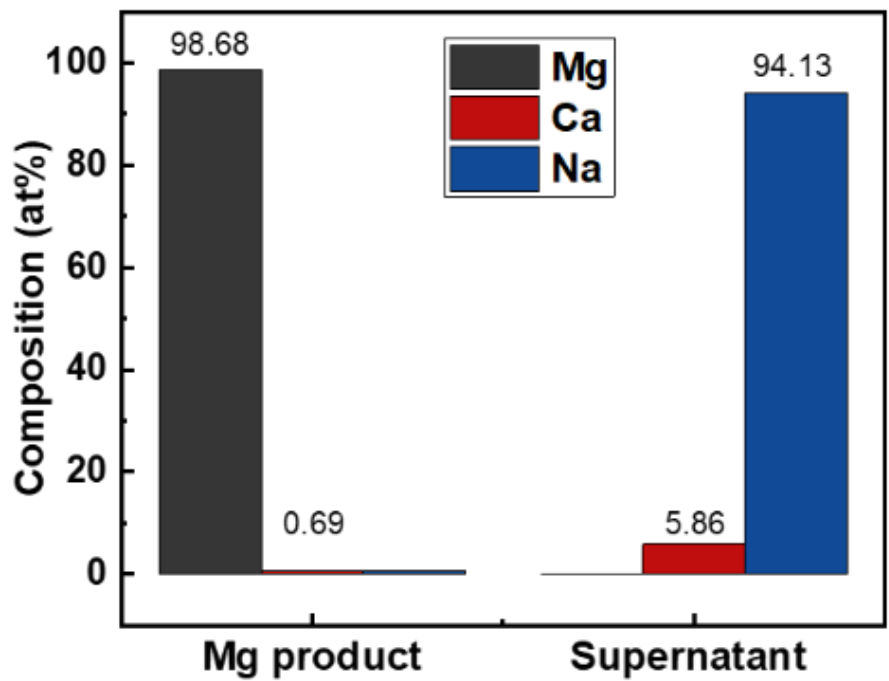

**Fig S3. Reproducibility of the Opentrons Separations Experiments**

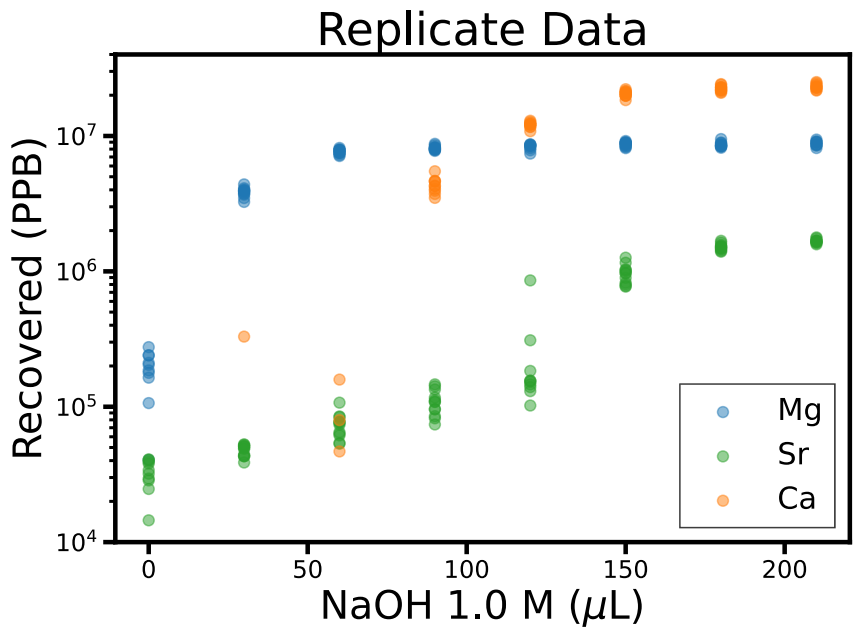

a.

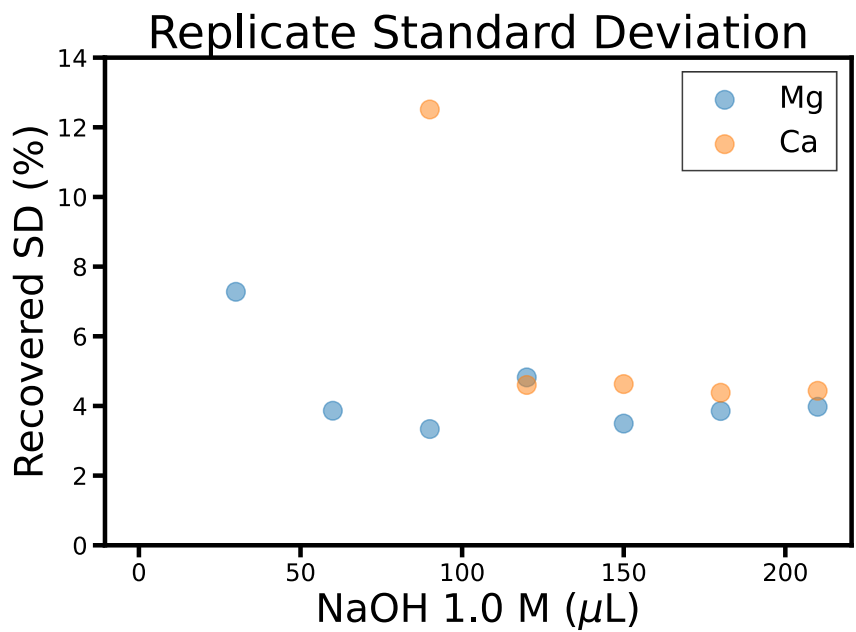

b.

Typical standard deviation of elements measured by ICP-MS was determined via 12 replicate experiments at each condition of the produced water precipitation. Higher standard deviations were observed prior to complete precipitation of an element and for trace elements.

**Figure S4. High-Throughput UV-Vis Spectra**

a.

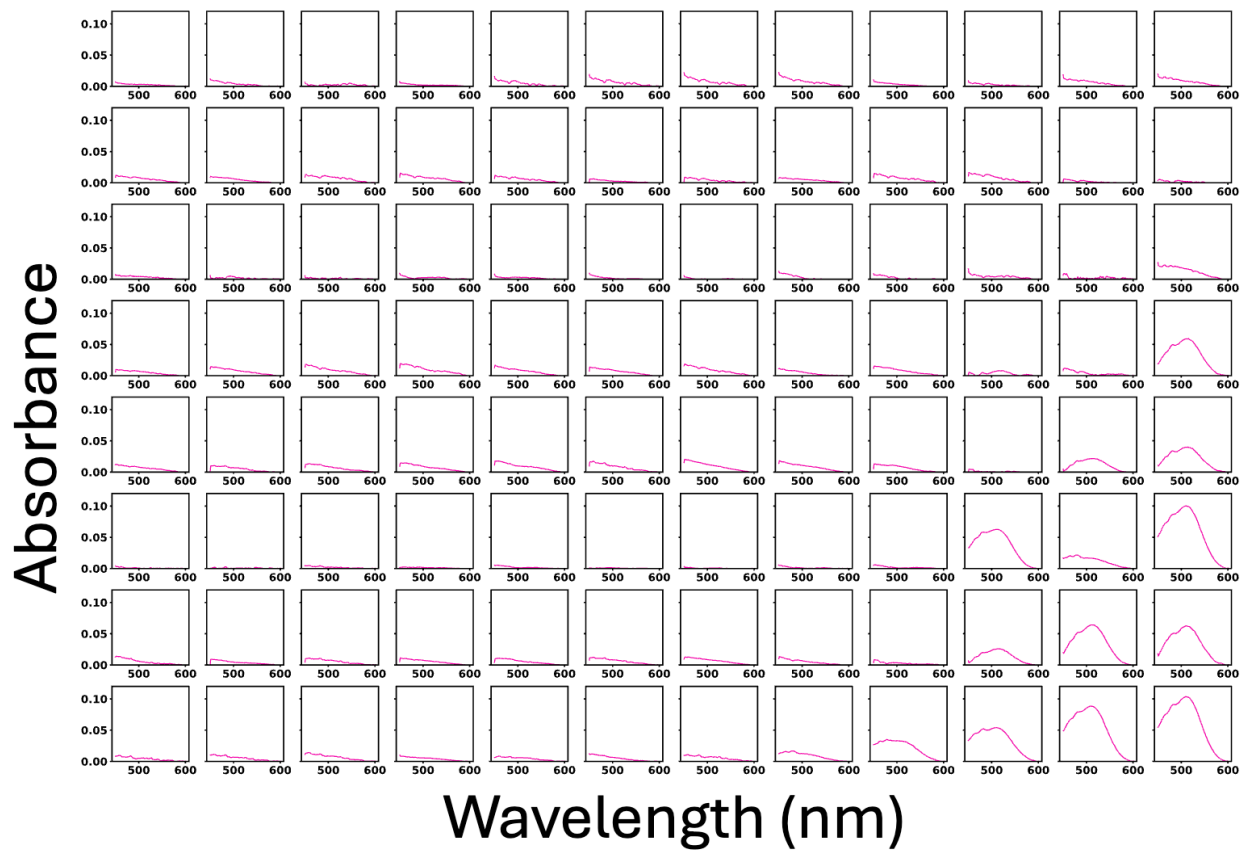

b.

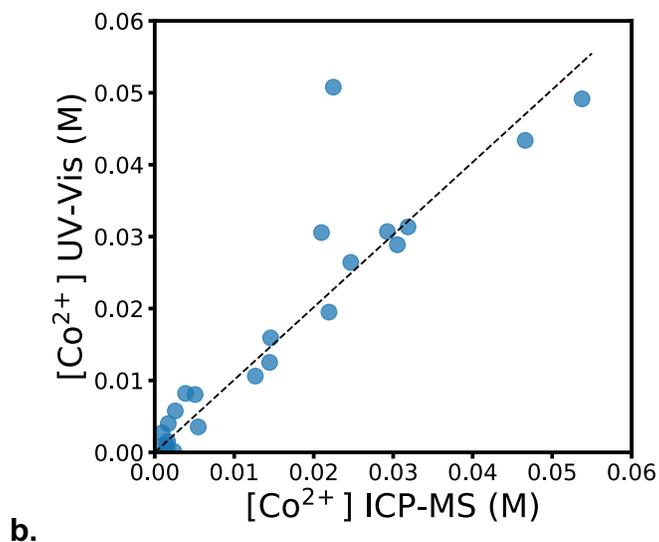

High throughput UV-Vis spectra were obtained using a Thermofisher Varioskan LUX microplate reader. SmCo digests were measured from 460 to 610 nm. The concentrations determined from this method were compared to values from ICP-MS and found to be in highly linear with slope = 0.97. These results support UV-Vis as a robust high throughput metric for automated experiments where UV-Vis signatures are available. Example data are shown for Round 2 of SmCo extraction.

**Fig S5. Comparison of Purity vs Yield**

For the experimental campaigns described in the text, purity of the desired element (Sm/Nd/Pr) was used and found to be a robust metric for the planning and Bayesian optimization agents. A more detailed technoeconomic metric considering factors such as the nonlinearity of purity and value is beyond the scope of this study, however would be possible with the current agentic capabilities and could lead to different optimizations.

a. NdFeB Round 2:

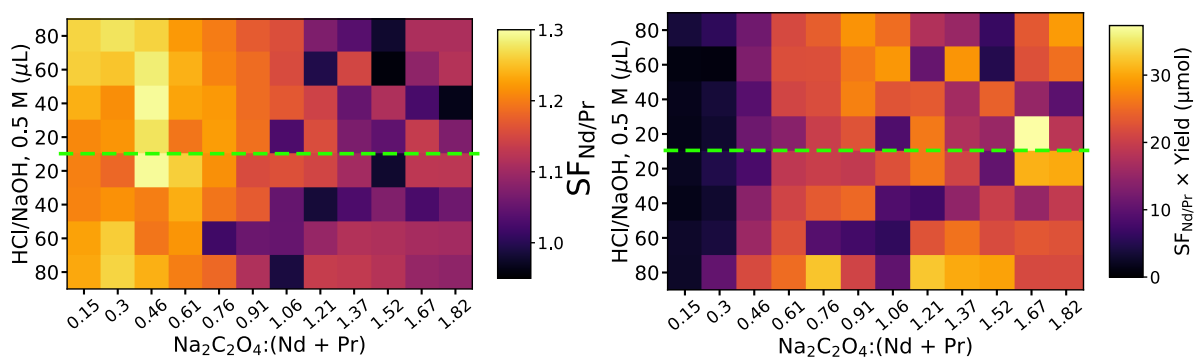

Within the current datasets we consider the product of separation factor and yield as a first order approximation of this. It is clearly seen that while the low-equivalent regime of added oxalate has the best Nd:Pr separation factor, the absolute yield and by extension value of extracted material is much lower.

b. SmCo Round 2:

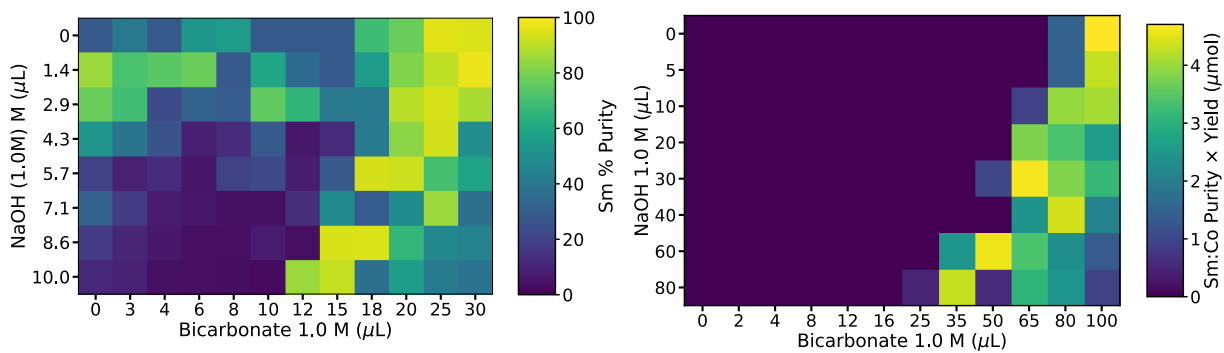

## Fig S6. Bayesian Optimization Acquisition Functions

SmCo Round 1 Results:

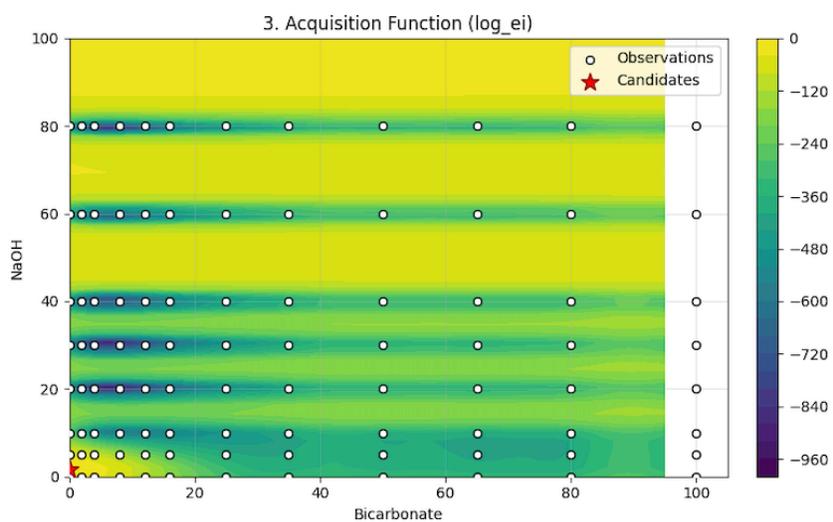

SmCo Round 2 Results:

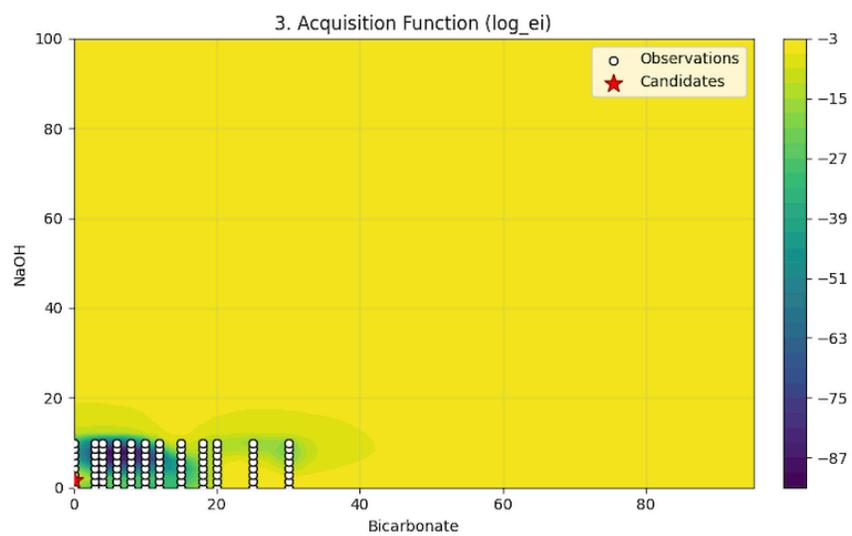

## Fig S7. SciLink Prompts

The following input prompts were used when generating experimental plans using SciLink's natural language interface. Full outputs are available in supplemental json files.

**Techno-economic Objective (Same for all cases):**
   "Using the DOE assessment report, the PWS database, and the provided criticality matrix image as context, analyze the ICP-MS results to determine which "measured critical materials show concentrations that might be economically interesting for recovery, considering their market value."

**Experiment Objective (Same for all round 1 cases):**
   "Based on the ICP-MS results and the prior TEA identifying valuable materials, a simple chemical process (like precipitation) to selectively recover the most promising material from the water sample. Use only reagents that are simple commodity chemicals. In your proposal specifically identify a range of conditions for concentrations, ratios, solubilities, or other variable for testing optimal recovery. Put these conditions in a table for a 96 well plate and using an opentrons. The experiment should cover a wide range of conditions which may span into the non-recovery regime, so as to provide the most data. Rather than measure pH estimate required amounts of acid or base. Prefer additional conditions spread over the columns and rows rather than replicate trials. Target a maximum 360 µL total volume, and prefer to use concentrations at most 1M. Provide corresponding Opentrons code, requirements for this code are in Opentrons_Setup.txt"

**Second Round NdFeB Experiment Objective:**
Here are the results of an experiment trying to preferentially precipitate and separate Nd from a stock solution made from NdFeB. The experiment used NaOH and NaOxalate in various amounts to see which did a better job of extracting Nd. The experiment done was by Opentrons and is in BFeNd.py. You should also observe the png graphed results. There are two subsquent tasks for the next experiment:
   1: Determine which method was superior overall, NaOH or NaOxalate and justify
   2: Create a plan for further purifying the Nd containing material, which also contains other trace elements
The requirements for experiment are as before:
 Propose a simple chemical process (like precipitation) to selectively recover the most promising material from the water sample. Use only reagents that are simple commodity chemicals. In your proposal specifically identify a range of conditions for concentrations, ratios, solubilities, or other variable for testing optimal recovery of the most valuable component(s). Arrange these conditions as a matrix for a 96 well plate and using an opentrons. The experiment should cover a wide range of conditions which may span into the non-recovery regime, so as to provide the most data. Rather than measure pH estimate required amounts of acid or base. Prefer additional conditions spread over the columns and rows rather than replicate trials. Target a maximum 400 µL total volume, and prefer to use concentrations at most 1M. Provide corresponding Opentrons code, ensure compliance of this code with descriptions in Opentrons_Setup.txt.

**Second Round SmCo BO-Informed Experiment Objective:**
Our goal is to create a second round of experiments to refine results of a precipitation experiment. We previously Dispensed 1.0M NaOH along the 8 rows of a 96 wellplate and 1.0M Bicarbonate along the 12 columns. Bayesian optimization has found the following values: Compose round 2 as another 96 plate with NaOH and Bicarbonate in rows and columns. Based on these values in batch_step_1.csv compose a range of conditions to span the 96 wellplate that reflect the bayesian values and clustering. Target a maximum 400 uL total volume, using 200 of starting material and use water for any missing volume below that 400. Provide corresponding Opentrons Code, requirements for this code are in Opentrons_Setup.txt and api context in opentrons_api.txt.

**Fig S8. Code Feedback Examples**

The natural language feedback feature can be used for both small adjustments and corrections or larger semantic changes to experimental design:

Feedback given for SmCo Round 2:
Example A.

```python
tiprack = protocol.load_labware('opentrons_flex_96_tiprack_1000ul', 'C2')

pipette.pick_up_tip()
```

Feedback:
"make sure to specify using the smaller 200 tips for NaOH and Bicarbonate since those values are small"

```python
tiprack_1000 = protocol.load_labware('opentrons_flex_96_tiprack_1000ul', 'C2')
tiprack_200 = protocol.load_labware('opentrons_flex_96_tiprack_200ul', 'C3')

pipette.pick_up_tip(tiprack_200)
```

Example B.

```python
        # Dispense to each well
for row_idx, row in enumerate(rows):
    for col_idx, col in enumerate(columns):
        well = plate[f'{row}{col}']
        naoh_vol = naoh_volumes[row_idx]
        bicarb_vol = bicarbonate_volumes[col_idx]
        water_vol = 200 - naoh_vol - bicarb_vol

        # Ensure water volume is not negative
        if water_vol < 0:
            water_vol = 0

        # Dispense leachate (200 µL)
        pipette.pick_up_tip()
        pipette.aspirate(200, leachate)
        pipette.dispense(200, well)
        pipette.drop_tip()

        # Dispense NaOH if volume > 0
        if naoh_vol > 0:
            pipette.pick_up_tip()
            pipette.aspirate(naoh_vol, naoh)
            pipette.dispense(naoh_vol, well)
            pipette.drop_tip()
```

```python
        # Dispense Bicarbonate if volume > 0
        if bicarb_vol > 0:
        pipette.pick_up_tip()
        pipette.aspirate(bicarb_vol, bicarbonate)
        pipette.dispense(bicarb_vol, well)
        pipette.drop_tip()

        # Dispense water if volume > 0
        if water_vol > 0:
        pipette.pick_up_tip()
        pipette.aspirate(water_vol, water)
        pipette.dispense(water_vol, well)
        pipette.drop_tip()
```

Feedback:
"change the looping to go in order: leachate, water, NaOH, bicarbonate. Finish each reagent for the entire plate then move to the next"

Revised Code Snippet:
```python
# Step 1: Dispense leachate (200 \u00b5L) to all wells using 1000 \u00b5L tips
    protocol.comment(\"Step 1: Dispensing leachate to all wells\")
    for row in rows:
        for col in columns:
            well = plate[f'{row}{col}']
            pipette.pick_up_tip(tiprack_1000)
            pipette.aspirate(LEACHATE_VOLUME, leachate)
            pipette.dispense(LEACHATE_VOLUME, well)
            pipette.drop_tip()

    # Step 2: Dispense water to all wells (where volume > 0) using 1000 \u00b5L tips
    protocol.comment(\"Step 2: Dispensing water to all wells\")
    for row in rows:
        for col in columns:
            well_name = f'{row}{col}'
            well = plate[well_name]
            water_vol = water_volumes[well_name]

            if water_vol > 0:
                pipette.pick_up_tip(tiprack_1000)
                pipette.aspirate(water_vol, water)
                pipette.dispense(water_vol, well)
                pipette.drop_tip()

    # Step 3: Dispense NaOH to all wells (where volume > 0) using 200 \u00b5L tips for accuracy
```

```python
        protocol.comment(\"Step 3: Dispensing NaOH to all wells (using 200 \u00b5L tips for small volumes)\")
        for row_idx, row in enumerate(rows):
            naoh_vol = naoh_volumes[row_idx]

            if naoh_vol > 0:
                for col in columns:
                    well = plate[f'{row}{col}']
                    pipette.pick_up_tip(tiprack_200)
                    pipette.aspirate(naoh_vol, naoh)
                    pipette.dispense(naoh_vol, well)
                    pipette.drop_tip()

        # Step 4: Dispense Bicarbonate to all wells (where volume > 0) using 200 \u00b5L tips for accuracy
        protocol.comment(\"Step 4: Dispensing bicarbonate to all wells (using 200 \u00b5L tips for small volumes)\")
        for col_idx, col in enumerate(columns):
            bicarb_vol = bicarbonate_volumes[col_idx]

            if bicarb_vol > 0:
                for row in rows:
                    well = plate[f'{row}{col}']
                    pipette.pick_up_tip(tiprack_200)
                    pipette.aspirate(bicarb_vol, bicarbonate)
                    pipette.dispense(bicarb_vol, well)
                    pipette.drop_tip()
```

## References:

(1) M. Douze et al., "THE FAISS LIBRARY," in IEEE Transactions on Big Data, doi: 10.1109/TBDATA.2025.361847.

(2) Lyman, J.; Palmer, G., Recycling of Rare Earths and Iron from NdFeB Magnet Scrap. *High Temperature Materials and Processes* **1993,** *11* (1-4), 175-188.

(3) Kumari, A.; Jha, M. K.; Pathak, D. D., An Innovative Environmental Process for the Treatment of Scrap Nd-Fe-B Magnets. *Journal of Environmental Management* **2020,** *273*, 111063.